\documentclass[seceq]{ptptex}
\usepackage{wrapft}
\usepackage{graphicx}

\newcommand{\ket}[1]{| {#1} \rangle}     
\newcommand{\rket}[1]{| {#1} )}     
\newcommand{\dbra}[1]{\langle {#1} |\!|}     
\newcommand{\dket}[1]{|\!| {#1} \rangle}     

\newcommand{\wtilde}[1]{\widetilde{#1}} 

\def\beq{\begin{eqnarray}}
\def\eeq{\end{eqnarray}}
\def\bsub{\begin{subequations}}
\def\esub{\end{subequations}}
\def\b{\begin{equation}}

\title{
Re-formation of Many-Quark Model with the $su(4)$-Algebraic Structure in the 
Schwinger Boson Realization
}
\subtitle{
Reconsideration in the Original Fermion Space
}    

\author{
Yasuhiko {\sc Tsue},$^{1}$ 
Constan\c{c}a {\sc Provid\^encia},$^{2}$ 
Jo\~ao da {\sc Provid\^encia}$^{2}$ and 
Masatoshi {\sc Yamamura}$^{3}$  
}

\inst{
$^{1}$Physics Division, Faculty of Science, Kochi University, Kochi 780-8520, 
Japan\\
$^{2}$Departamento de F\'{i}sica, Universidade de Coimbra, 3004-516 Coimbra, 
Portugal\\
$^{3}$Department of Pure and Applied Physics, 
Faculty of Engineering Science,\\
Kansai University, Suita 564-8680, Japan
\\

}

\recdate{
\today
}

\abst{
The energy eigenstates for the $su(4)$-algebraic model for many-quark system 
obtained by the present authors in the Schwinger boson space are 
reconsidered in the original fermion space. 
Through this task, the structures of the single-quark and the quark-triplet are clarified. 
Variations of the $su(4)$-generators which have been adopted by the present authors are discussed. 
}

\begin{document}

\maketitle

\section{Introduction}

The $su(4)$algebraic model for many-quark system may be an attractive model. 
With the aid of this model, we are able to obtain a schematic understanding of 
not only the quark-triplet but also the quark-pair phase. 
In addition, if we are interested in the single-quark phase, this model enables us 
to investigate this phase including the possibility of its existence. 
Investigation of many-quark system based on the $su(4)$-algebra traces back to 1983. 
In this year, Petry et al. proposed an interesting model which has been called the 
Bonn model.\cite{1} 
This model leads us to the quark-triplet which may be identified as ``nucleon" 
in the spherical $j$-$j$ coupling shell model including $\Delta$-particle. 
In succession, aiming at realistic application, some works have been reported.\cite{2} 
The Bonn model is a kind of effective models based on the quark-pairing interaction. 
Therefore, in the frame of the Bonn model, we may expect to describe the 
quark-pairing phase, namely, color superconducting phase.\cite{3} 
Recently, with the aim of investigating the quark-triplet and the quark-pair phase 
in a unified scheme, the present authors have been concerned with the 
Bonn model.\cite{4,5} 
Naturally, preserving the basic aspects of this model, 
we modified it so as to be able to correct certain insufficient parts. 
Our description is based on the Schwinger boson representation proposed by the 
present authors (M.Y.) with Kuriyama and Kunihiro.\cite{6} 
The energy eigenvalues and their eigenstates are exactly obtained and we gave various analysis for the results. 
However, as will be later mentioned, 
practically, we did not consider the color-symmetric nature of the model explicitly.

Under the above-mentioned circumstances, very recently, the present authors have presented three 
papers,\cite{7} which will be referred to as (I) collectively. 
In the same idea as that of Refs.\citen{4} and \citen{5}, 
the $su(4)$-algebra was treated in the Schwinger boson space. 
Of course, we took care of the correspondence to the original fermion 
space. 
Usually, the Lie algebraic approach to many-body theory starts in a 
chosen minimum weight state. 
We also follow this idea. 
The starting form of our $su(4)$-algebraic model is color-symmetric 
and involves six types of minimum weight states. 
The eigenvalues of the Casimir operator are all identical to one another. 
In these six, we choose a certain one. 
As a natural consequence, the basic equations obtained under the chosen minimum 
weight state violate the color-symmetry and the eigenstates of the 
color-symmetric Hamiltonian as a whole are not color-singlets. 
In Refs.\citen{4} and \citen{5}, 
we did not consider this property explicitly. 
In (I), we gave a certain reasonable interpretation for this problem and showed that 
the energy eigenvalues are of the same forms as those obtained in Refs.\citen{4} and 
\citen{5}. 
On the basis of these results, we made detailed analyses for the ground-state energies with 
positive results. 
For example, the treatment in (I) gives the following conclusion: 
In the low and the high density region, the quark-triplet and the 
quark-pair phase are dominant, respectively. 
The above is consistent to the common understanding. 
The description of (I) is based on a specific bilinear form of the 
quark operators for the $su(4)$-generators and concrete 
analysis is performed not in the original fermion space, but in the Schwinger boson space. 
Therefore, for instance, in the framework of (I), it is an open question to 
identify the quark-triplet obtained in (I) with the ``nucleon" in the spherical 
$j$-$j$ coupling shell model which characterizes the Bonn model and, 
further, the structure of the single-quark operator is not 
clarified in relation to the original fermion operator.

The first aim of this paper is to transcribe the eigenstates of the Hamiltonian 
obtained in the Schwinger boson space into the corresponding eigenstates in the 
original fermion space. 
The idea for the transcription is simple. 
As was shown in (I), the eigenstates in the Schwinger boson space are derived 
by operating certain state-generating operators on the 
minimum weight states. 
The state-generating operators are certain functions of the $su(4)$-generators. 
In Ref.\citen{5}, we presented other forms by performing this operation explicitly. 
Through this process, besides the $su(4)$-generators as the quark-pair 
creation operators, we could derive the explicit forms of the 
single-quark and the quark-triplet creation operators in the Schwinger boson 
space, namely, the building blocks for constructing the eigenstates. 
Therefore, it may be enough for our task to find the minimum weight states and the 
correspondences of the state-generating operators in the original fermion space. 
Through this transcription, we can learn that the single-quark operators 
are the quark operator themselves and also the structure 
of the quark-triplets are clarified. 
In other words, the building blocks for constructing the eigenstates are given.

The second aim is to present various variations of the $su(4)$-generators in the 
original fermion space obtained in (I). 
These are expressed in terms of the bilinear forms of the 
quark creation and annihilation operators. 
Each quark operator is specified by a set of quark numbers for 
the single-particle state besides the color quantum number. 
Coefficients of the bilinear forms are determined so as to lead to the 
$su(4)$-algebra. 
Therefore, the coefficients depend on the quantum numbers specifying the 
single-particle states. 
Following the choice of the quantum numbers, 
various $su(4)$-algebraic models appear. 
In (I), we treated the following case: 
The total number of the single-particle states, each of which is specified by 
one quantum number, is even. 
For our idea for describing the $su(4)$-algebraic model, this case may be very instructive. 
In this paper, after giving a general form, we will show 
mainly three variations of the expressions for the $su(4)$-generators. 
The first and the second variation are for the hadron and nuclear physics. 
In particular, the relation between the quark-triplet and ``nucleon" in the 
shell model is discussed. 
The third is for the atomic physics:
The model for a trapped three color atom 
gas is presented by Errea et al.\cite{8}

In next two sections, some aspects of the model treated in (I) are recapitulated. 
In \S 2, mainly, the outline of the model is given both in the fermion space 
and in its corresponding Schwinger boson space. 
In \S 3, the eigenstates of the Hamiltonian are given in the Schwinger boson space. 
The minimum weight state and the state-generating operators are introduced. 
Section 4 is devoted to presenting the explicit 
forms of the eigenstates in the original fermion space. 
In \S 5, the general expression of the $su(4)$-generators is presented in terms of 
covering the form recapitulated in \S 2. 
Further, the eigenstates are given for the general case. 
In \S 6, the variations of the model given in \S 2 are discussed. 
The last section is devoted to a brief summary. 
In Appendix, some mathematical formulae are given.

\section{The $su(4)$-algebraic many-quark model and its Schwinger boson realization}

In this section, we will recapitulate some aspects of our model by rearranging the results 
reported in our several papers,\cite{4,5,7}
mainly in (I). 
This model is formulated in terms of the generators of the $su(4)$-algebra 
constructed by the bilinear forms of the quark operators. 
The color quantum number is denoted by $i=1$, 2 and 3. 
Each color state has the degeneracy $(2j_s+1)$. 
Here, $j_s$ denotes a half-integer. 
For the present, the degrees of freedom related to isospin are 
ignored. 
Therefore, any single-particle state is specified as $(i, m)$ with 
$m=\pm 1/2$, $\pm 3/2$, $\cdots$, $\pm j_s$. 
It should be noted that the above specification for the single-particle 
state does not always mean the spherical many-fermion system. 
In \S 5, we will reconsider this statement. 
Creation and annihilation operators are denoted as 
${\tilde c}_{im}^*$ and ${\tilde c}_{im}$, respectively.

We introduce the following fifteen operators which are investigated in (I): 
\bsub\label{2-1}
\beq
& &{\wtilde S}^1=\sum_m{\tilde c}_{2m}^*{\tilde c}_{3{\tilde m}}^*\ , \quad
{\wtilde S}^2=\sum_m{\tilde c}_{3m}^*{\tilde c}_{1{\tilde m}}^*\ , \quad
{\wtilde S}^3=\sum_m{\tilde c}_{1m}^*{\tilde c}_{2{\tilde m}}^*\ , \nonumber\\
& &{\wtilde S}_1=\left({\wtilde S}^1\right)^*\ , \qquad
{\wtilde S}_2=\left({\wtilde S}^2\right)^*\ , \qquad
{\wtilde S}_3=\left({\wtilde S}^3\right)^*\ , 
\label{2-1a}\\
& &{\wtilde S}_1^2=-\sum_m{\tilde c}_{2m}^*{\tilde c}_{1m}\ , \quad
{\wtilde S}_1^3=-\sum_m{\tilde c}_{3m}^*{\tilde c}_{1m}\ , \quad
{\wtilde S}_2^3=-\sum_m{\tilde c}_{3m}^*{\tilde c}_{2m}\ , \nonumber\\
& &{\wtilde S}_2^1=\left({\wtilde S}_1^2\right)^*\ , \qquad
{\wtilde S}_3^1=\left({\wtilde S}_1^3\right)^*\ , \qquad
{\wtilde S}_3^2=\left({\wtilde S}_2^3\right)^*\ , 
\label{2-1b}\\
& &{\wtilde S}_1^1=\sum_m({\tilde c}_{2m}^*{\tilde c}_{2m}+
{\tilde c}_{3m}^*{\tilde c}_{3m}-1) \ , \quad
{\wtilde S}_2^2=\sum_m({\tilde c}_{3m}^*{\tilde c}_{3m}+
{\tilde c}_{1m}^*{\tilde c}_{1m}-1) \ , \nonumber\\
& &{\wtilde S}_3^3=\sum_m({\tilde c}_{1m}^*{\tilde c}_{1m}+
{\tilde c}_{2m}^*{\tilde c}_{2m}-1) \ . 
\label{2-1c}
\eeq
\esub 
Here, we use the notation ${\tilde m}$ for $-m$. 
The operators (\ref{2-1}) form the $su(4)$-algebra:
\bsub\label{2-2}
\beq
& &[\ {\wtilde S}^i\ , \ {\wtilde S}^j\ ]=0\ , \qquad
[\ {\wtilde S}^i\ , \ {\wtilde S}_j\ ]={\wtilde S}_i^j \ , \nonumber\\
& &[\ {\wtilde S}_i^j\ , \ {\wtilde S}^k\ ]=\delta_{ij}{\wtilde S}^k +\delta_{jk}{\wtilde S}^i\ , \quad
[\ {\wtilde S}_i^j\ , \ {\wtilde S}_l^k\ ]=\delta_{jl}{\wtilde S}_i^k -\delta_{ik}{\wtilde S}_l^j \ , 
\label{2-2a}
\eeq
The Casimir operator ${\wtilde {\mib P}}^2$ is expressed as 
\beq
{\wtilde {\mib P}}^2&=&
\sum_i\left({\wtilde S}_i{\wtilde S}^i+{\wtilde S}^i{\wtilde S}_i\right)
+\sum_{i\neq j}{\wtilde S}_j^i{\wtilde S}_i^j \nonumber\\
& &+\frac{1}{4}\left[
\left({\wtilde S}_2^2-{\wtilde S}_3^3\right)^2+\left({\wtilde S}_3^3-{\wtilde S}_1^1\right)^2
+\left({\wtilde S}_1^1-{\wtilde S}_2^2\right)^2\right]\ .
\label{2-2b}
\eeq
\esub
As a subalgebra, the $su(4)$-algebra contains the $su(3)$-algebra. 
For example, the following set forms the $su(3)$-algebra:
\bsub\label{2-3}
\beq
{\wtilde S}_1^2\ , \ \ {\wtilde S}_2^1\ , \ \ 
{\wtilde S}_1^3\ , \ \ {\wtilde S}_3^1\ , \ \ 
{\wtilde S}_2^3\ , \ \ {\wtilde S}_3^2\ , \ \ 
\frac{1}{2}\left({\wtilde S}_2^2-{\wtilde S}_3^3\right)\ , \ \ 
{\wtilde S}_1^1-\frac{1}{2}\left({\wtilde S}_2^2+{\wtilde S}_3^3\right)\ .
\label{2-3a}
\eeq
The Casimir operator ${\wtilde {\mib Q}}^2$ is given in the form 
\beq
{\wtilde {\mib Q}}^2&=&
\sum_{i\neq j}{\wtilde S}_j^i{\wtilde S}_i^j +\frac{1}{3}\left[
\left({\wtilde S}_2^2-{\wtilde S}_3^3\right)^2+\left({\wtilde S}_3^3-{\wtilde S}_1^1\right)^2
+\left({\wtilde S}_1^1-{\wtilde S}_2^2\right)^2\right] \nonumber\\
&=&\sum_{i\neq j}{\wtilde S}_j^i{\wtilde S}_i^j+
2\left[\frac{1}{2}\left({\wtilde S}_2^2-{\wtilde S}_3^3\right)\right]^2
+\frac{2}{3}\left[{\wtilde S}_1^1-
\frac{1}{2}\left({\wtilde S}_2^2+{\wtilde S}_3^3\right)\right]^2\ . 
\label{2-3b}
\eeq
\esub
We can see that ${\wtilde {\mib P}}^2$ and ${\wtilde {\mib Q}}^2$ are color-symmetric. 
By permutations from $(1,2,3)$ for the color quantum numbers to others, 
for example, such as $(2,3,1)$, we have six cases including the identical permutation. 
In order to formulate the present model in terms of the color-symmetric form, 
it may be necessary to treat the above six cases 
on an equal footing. 
It was stressed in (I). 
At the moment, 
our discussion is restricted to the case (\ref{2-3a}). 
In addition, the operator, which is linearly independent of 
$({\wtilde S}_2^2-{\wtilde S}_3^3)/2$ and ${\wtilde S}_1^1-({\wtilde S}_2^2+{\wtilde S}_3^3)/2$, 
is introduced:
\beq\label{2-4}
{\wtilde P}_0=\frac{1}{2}\left({\wtilde S}_1^1+{\wtilde S}_2^2+{\wtilde S}_3^3\right)\ .
\eeq
The operator ${\wtilde P}_0$ is color-symmetric and satisfies 
\beq\label{2-5}
[\ {\wtilde S}_i^j \ , \ {\wtilde P}_0\ ]=0\ . \qquad
(i,\ j=1,\ 2,\ 3)
\eeq

The Hamiltonian adopted in this model is expressed in the form 
\beq\label{2-6}
{\wtilde H}_m={\wtilde H}+{\tilde \chi}{\wtilde {\mib Q}}^2 \ , \qquad
{\wtilde H}=-\sum_i{\wtilde S}^i{\wtilde S}_i \ . 
\eeq
If ${\tilde \chi}=0$, ${\wtilde H}_m$ reduces to ${\wtilde H}$. 
It is similar to the form originally adopted in the Bonn model. 
The Hamiltonian ${\wtilde H}$ is characterized by the relation 
\beq\label{2-7}
[\ {\wtilde S}_i^j \ , \ {\wtilde H}\ ]=0 \ .
\eeq
In order to explain certain properties of a many-quark system which 
cannot be explained in the frame of ${\wtilde H}$, the present authors 
modified ${\wtilde H}$ to ${\wtilde H}_m$ by adding the term 
${\tilde \chi}{\wtilde {\mib Q}}^2$ to ${\wtilde H}$ in (I). 
We are interested in the change of the energy eigenvalues of ${\wtilde H}_m$ 
from those of ${\wtilde H}$ without any change for the energy eigenstates. 
Therefore, we require the condition 
\beq\label{2-8}
[\ {\wtilde S}_i^j \ , \ {\tilde \chi}\ ]=0\ , \quad {\rm i.e.}\quad
[\ {\wtilde S}_i^j \ , {\wtilde H}_m\ ]=0 \ .
\eeq
The above is the outline of the framework of our model.

In (I), we have investigated in detail the above $su(4)$-model in the framework 
of the Schwinger boson realization. 
By introducing eight kinds of boson operators $({\hat a}, {\hat a}^*)$, 
$({\hat b}, {\hat b}^*)$, $({\hat a}_i, {\hat a}_i^*)$ and $({\hat b}_i, {\hat b}_i^*)$ 
$(i=1,2,3)$, the $su(4)$-algebra can be formulated as follows: 
\beq\label{2-9}
& &{\wtilde S}^i\rightarrow {\hat S}^i={\hat a}_i^*{\hat b}-{\hat a}^*{\hat b}_i \ , \quad
{\wtilde S}_i\rightarrow {\hat S}_i={\hat b}^*{\hat a}_i-{\hat b}_i^*{\hat a} \ , \nonumber\\
& &{\wtilde S}_i^j\rightarrow {\hat S}_i^j=({\hat a}_i^*{\hat a}_j-{\hat b}_j^*{\hat b}_i)
+\delta_{ij}({\hat a}^*{\hat a}-{\hat b}^*{\hat b}) \ .
\eeq  
Associating with the above $su(4)$-algebra, we can define the $su(1,1)$-algebra: 
\beq\label{2-10}
& &{\hat T}_+={\hat a}^*{\hat b}^*+\sum_i{\hat a}_i^*{\hat b}_i^* \ , \qquad
{\hat T}_-={\hat b}{\hat a}+\sum_i{\hat b}_i{\hat a}_i \ , \nonumber\\
& &{\hat T}_0=\frac{1}{2}({\hat a}^*{\hat a}+{\hat b}^*{\hat b})+
\frac{1}{2}\sum_i({\hat a}_i^*{\hat a}_i+{\hat b}_i^*{\hat b}_i)+2 \ .
\eeq
The commutation relations for ${\hat T}_{\pm, 0}$ and the Casimir operator 
${\hat {\mib T}}^2$ are given in the form 
\bsub\label{2-11}
\beq
& &[\ {\hat T}_+\ , \ {\hat T}_- \ ]=-2{\hat T}_0\ , \qquad
[\ {\hat T}_0\ , \ {\hat T}_{\pm} \ ]=\pm {\hat T}_{\pm}\ , 
\label{2-11a}\\
& &{\hat {\mib T}}^2=-\frac{1}{2}\left({\hat T}_-{\hat T}_++{\hat T}_+{\hat T}_-\right)+{\hat T}_0^2\ . 
\label{2-11b}
\eeq
\esub
It may be important to see the relation 
\beq\label{2-12}
[\ {\rm any\ of}\ ({\hat T}_{\pm,0})\ , \ {\rm any\ of}\ ({\hat S}^i,\ {\hat S}_i,\ {\hat S}_i^j)\ ]=0 \ .
\eeq

By replacing ${\wtilde S}^i$ etc. and ${\tilde \chi}$ in ${\wtilde H}_m$ with ${\hat S}^i$ etc. and 
${\hat \chi}$, respectively, we obtain the Hamiltonian in the Schwinger boson space, 
which is denoted as ${\hat H}_m$. 
Of course, if ${\tilde \chi}$ is $c$-number, i.e., 
${\tilde \chi}=\chi$, ${\hat \chi}$ should be also $c$-number, i.e., ${\hat \chi}=\chi$ and 
the relation (\ref{2-12}) leads us to 
\beq\label{2-13}
[\ {\hat T}_{\pm,0}\ , \ {\hat H}_m\ ]=0 \ .
\eeq
As a concrete example, we have shown the form of ${\hat \chi}$ as a $q$-number in (I):
${\hat \chi}$ is a certain function of ${\hat {\mib T}}^2$ and ${\hat P}_0$. 
The operator ${\hat P}_0$ is obtained by replacing ${\wtilde S}_i^i$ with 
${\hat S}_i^i$ $(i=1,2,3)$ in ${\wtilde P}_0$ defined in the relation (\ref{2-4}). 
In this case, we have 
\beq\label{2-14}
[\ {\hat T}_{\pm,0} \ , \ {\hat \chi}\ ]=0 \ .
\eeq
Then, under this condition, the relation (\ref{2-13}) is realized. 
Since ${\hat \chi}$ is regarded as a function of ${\hat {\mib T}}^2$ and ${\hat P}_0$, we have 
\beq\label{2-15}
[\ {\hat S}_i^j \ , \ {\hat \chi}\ ]=0 \ , \quad {\rm i.e.,}\quad 
[\ {\hat S}_i^j\ , \ {\hat H}_m\ ]=0 \ . 
\eeq
Under an appropriate choice of ${\hat \chi}$, our model interprets one of the important features of 
many-quark system: 
in the low and in the high density region, the quark-triplet and the quark-pair phase are dominant, respectively.

The above is an outline of our model in the Schwinger boson 
realization. 
Of course, it has been already reported in several papers\cite{4,5} including 
(I). 
In this connection, we must comment the following: 
The form (\ref{2-9}) is by no means unique.  
For example, the simplest case is presented by the form 
${\hat S}^i={\hat a}_i^*{\hat b}$, ${\hat S}_i={\hat b}^*{\hat a}_i$ and 
${\hat S}_i^j={\hat a}_i^*{\hat a}_j-\delta_{ij}{\hat b}^*{\hat b}$. 
However, the quark-triplet cannot be treated by this representation.

\section{The energy eigenstates in the Schwinger boson space}

Our next task is to recapitulate the energy eigenstates of ${\hat H}_m$. 
First, we note that our Schwinger boson space is spanned by eight kinds of 
boson operators and, then, the orthogonal set is 
specified by eight quantum numbers. 
The relations (\ref{2-13}) and (\ref{2-14}) suggest that, in order to get 
the energy eigenvalues, it may be enough to search the minimum weight state 
$\ket{M_1}$ for the $su(1,1)$- and the $su(3)$-algebra: 
\bsub\label{3-1}
\beq
& &{\hat T}_-\ket{M_1}=0 \ , \qquad {\hat T}_0\ket{M_1}=(\sigma+2)\ket{M_1} \ , 
\nonumber\\
& &{\hat S}_2^1\ket{M_1}={\hat S}_3^1\ket{M_1}={\hat S}_3^2\ket{M_1}=0 \ , 
\label{3-1a}\\
& &\frac{1}{2}\left({\hat S}_2^2-{\hat S}_3^3\right)\ket{M_1}=-\lambda\ket{M_1} \ , \nonumber\\
& &\left[{\hat S}_1^1-\frac{1}{2}\left({\hat S}_2^2+{\hat S}_3^3\right)\right]\ket{M_1}
=-(2(\sigma-\sigma_0)+(\lambda-2\rho))\ket{M_1} \ , 
\label{3-1b}\\
& &{\hat P}_0\ket{M_1}=-(4(\lambda+\rho)-(\sigma+2\sigma_0))\ket{M_1} \ . 
\label{3-1c}
\eeq
\esub
Obviously, $\ket{M_1}$ is expressed in terms of the four quantum numbers: 
$\lambda$, $\rho$, $\sigma_0$ and $\sigma$. 
In (I), as the state $\ket{M_1}$ satisfying the relation (\ref{3-1}), we presented the form 
\beq
& &\ket{M_1}=\dket{\lambda\rho\sigma_0\sigma}=
\left({\hat S}^3\right)^{2\lambda}\left({\hat S}^4(1)\right)^{2\rho}({\hat b}_1^*)^{2(\sigma-\sigma_0)}
({\hat b}^*)^{2\sigma_0}\ket{0} \ , 
\label{3-2}\\
& &{\hat S}^4(1)=-{\hat S}^1\left(
{\hat S}_1^1-\frac{1}{2}\left({\hat S}_2^2+{\hat S}_3^3\right)\right)
-{\hat S}^2{\hat S}_1^2-{\hat S}^3{\hat S}_1^3 \ . 
\label{3-3}
\eeq
Here, in (I), we used ${\hat S}^4$ which is identical to $-{\hat S}^4(1)$. 
On the other hand, we presented another form 
\beq\label{3-4}
\ket{M_1}=\dket{lsrw}=({\hat S}^3)^{2l}
({\hat q}^1)^{2s}({\hat B}^*)^{2r}({\hat b}^*)^{2w}\ket{0} \ . 
\eeq
It has been proved in Ref.\citen{5} that both are equivalent to each other 
through the relation 
\beq\label{3-5}
l=\lambda\ , \qquad
s=\sigma-\sigma_0-\rho\ , \qquad r=\rho \ , \qquad w=\sigma \ . 
\eeq
The operators ${\hat q}^k$ and ${\hat B}^*$ are defined as 
\beq\label{3-6}
{\hat q}^k={\hat b}_k^*{\hat b}-{\hat a}^*{\hat a}_k\ , \qquad
{\hat B}^*=\sum_{i=1}^3 {\hat S}^i{\hat q}^i\ .
\eeq
The operators ${\hat S}^3$, ${\hat q}^1$ and ${\hat B}^*$ indicate the creation operators 
of the quark-pair, the single-quark and the quark-triplet, respectively, 
that is, the building blocks for the state $\dket{lsrw}$. 
They carry two, one and three quarks, respectively. 
Further, we have the relation 
\bsub\label{3-7}
\beq
& &[\ {\hat S}_i^j\ , \ {\hat S}^k\ ]=\delta_{ij}{\hat S}^k + \delta_{jk}{\hat S}^i \ , 
\label{3-7a}\\
& &[\ {\hat S}_i^j\ , \ {\hat q}^k\ ]=\delta_{ij}{\hat q}^k - \delta_{ik}{\hat q}^j \ , 
\label{3-7b}\\
& &[\ {\hat S}_i^j\ , \ {\hat B}^*\ ]=2\delta_{ij}{\hat B}^* \ . 
\label{3-7c}
\eeq
\esub
For the sake of the terms $\delta_{jk}{\hat S}^i$ and $\delta_{ik}{\hat q}^j$, 
${\hat S}^3$ and ${\hat q}^1$ are not color-singlet, but ${\hat B}^*$ is color-singlet. 
Therefore, we conclude that in spite of the eigenstate of ${\hat H}_m$, the state (\ref{3-4}) is not 
color-singlet.

Up to the present, our treatment was based on the $su(3)$-algebra shown in the form 
(\ref{2-3}). 
In \S 2, we mentioned that in order to guarantee the color-singlet property, we must take into account 
the other five forms obtained from the form (\ref{2-3}) by the permutation 
for (1, 2, 3). 
It can be seen that the state $\dket{lsrw}$ depends on the color quantum number 1 and 3 and, then, we use the notation 
$\dket{123;slrw}$ for $\dket{lsrw}$. 
We have two forms $\dket{231;slrw}$ and $\dket{312;slrw}$. 
Actually, the above three forms may give the idea of treating our model in the color-symmetric form. 
In this case, we can find the state $\dket{cs;slrw}$ in the form 
\beq\label{3-8}
\dket{cs;slrw}=\frac{1}{\sqrt{3}}\left(
\dket{123;slrw}+\dket{231;slrw}+\dket{312;slrw}\right) \ . 
\eeq 
Of course, $\dket{123;slrw}$ etc. are normalized. 
The state (\ref{3-8}) gives us the relations 
\beq
\dbra{cs;slrw}{\hat S}_i^j\dket{cs;slrw}&=&0\quad {\rm for}\quad i\neq j \ , 
\label{3-9}\\
\dbra{cs;slrw}{\hat S}_1^1\dket{cs;slrw}
&=&\dbra{cs;slrw}{\hat S}_2^2\dket{cs;slrw}
=\dbra{cs;slrw}{\hat S}_3^3\dket{cs;slrw} \nonumber\\
&=&\frac{4}{3}(s+2l)+2(2r-w)\ . 
\label{3-10}
\eeq
With the use of the relation (\ref{2-1c}) with the result (\ref{3-10}), we can show the result 
\beq\label{3-11}
\dbra{cs;slrw}{\hat N}_1\dket{cs;slrw}
&=&\dbra{cs;slrw}{\hat N}_2\dket{cs;slrw}
=\dbra{cs;slrw}{\hat N}_3\dket{cs;slrw}\nonumber\\
&=&\frac{N}{3}\ .
\eeq
Here, ${\hat N}_i$ denotes the quark-number operator of the color $i$, in the Schwinger boson 
realization:
\beq\label{3-12}
{\hat N}_i&=&
\frac{1}{2}(2j_s+1)+\frac{1}{2}\sum_j{\hat S}_j^j-{\hat S}_i^i \nonumber\\
&=&\frac{1}{2}(2j_s+1)+\frac{1}{2}({\hat a}^*{\hat a}-{\hat b}^*{\hat b})
+\frac{1}{2}\sum_j({\hat a}_j^*{\hat a}_j-{\hat b}_j^*{\hat b}_j)
-({\hat a}_i^*{\hat a}_i-{\hat b}_i^*{\hat b}_i)\ . 
\eeq
On the average, the color-singlet property of the state $\dket{cs;slrw}$ is 
guaranteed. 
The above is the outline of the energy eigenstates of our model in the Schwinger 
boson space which has been already reported in (I).

\section{The energy eigenstates in the original fermion space}

The main aim of this section is to investigate the explicit form of the energy 
eigenstates in the original fermion space. 
If we are interested only in obtaining the energy eigenvalues, it may be 
enough to treat the present model in the Schwinger boson realization. 
In fact, we have presented various aspects of the energy eigenvalues in the 
Schwinger boson space in (I). 
But, if we are interested also obtaining the energy eigenstates in terms of the constituents, 
we must turn back to the original fermion space. 
Through this task, we can complete our investigation of the $su(4)$-algebraic many-quark model.

First, we reinvestigate the state $\dket{\lambda\rho\sigma_0\sigma}$. 
This is obtained by operating with\break 
$({\hat S}^3)^{2\lambda}({\hat S}^4(1))^{2\rho}$ on the 
state $\ket{m_1}=\dket{\lambda=0,\rho=0,\sigma_0\sigma}$. 
In the above sense, ${\hat S}^3$ and ${\hat S}^4(1)$ can be regarded as the state-generating 
operators. 
Then, if we can find $\rket{m_1}$, which corresponds to $\ket{m_1}$, in the fermion space, the state we are looking for 
is expressed in the form $({\wtilde S}^3)^{2\lambda}({\wtilde S}^4(1))^{2\rho}\rket{m_1}$. 
Here, ${\wtilde S}^4(1)$ is obtained by replacing ${\hat S}^1$ etc. with ${\wtilde S}^1$ etc. in the 
form (\ref{3-3}). 
The state $\ket{m_1}$ is written down as
\beq\label{4-1}
\ket{m_1}=({\hat b}_1^*)^{2(\sigma-\sigma_0)}({\hat b}^*)^{2\sigma_0}\ket{0} \ . \qquad
(\sigma \geq \sigma_0)
\eeq
Then, we can specify the conditions characterizing $\ket{m_1}$ as follows: 
\bsub\label{4-2}
\beq
& &{\hat S}_1\ket{m_1}={\hat S}_2\ket{m_1}={\hat S}_3\ket{m_1}=0 \ , 
\label{4-2a}\\
& &{\hat S}_2^1\ket{m_1}={\hat S}_3^1\ket{m_1}={\hat S}_3^2\ket{m_1}=0 \ , 
\label{4-2b}\\
& &{\hat S}_1^1\ket{m_1}=-2\sigma\ket{m_1}\ , \qquad
{\hat S}_2^2\ket{m_1}=-2\sigma_0\ket{m_1}\ , \qquad
{\hat S}_3^3\ket{m_1}=-2\sigma'_0\ket{m_1} \ , \quad
\label{4-2c}\\
& &\sigma'_0=\sigma_0 \ . 
\label{4-2d}
\eeq
\esub
The above is nothing but the conditions to determine the minimum weight 
state for the $su(4)$-algebra in the case $\sigma'_0=\sigma_0$. 
We can formulate the Schwinger boson representation for the 
$\sigma'_0\neq \sigma_0$. 
But, this case may be unsuitable for the Schwinger boson representation for 
our present fermion model. 
This point has been discussed in Ref.\citen{4}.

Under the above argument, we set up the following conditions for the minimum weight state $\rket{m_1}$: 
\bsub\label{4-3}
\beq
& &{\wtilde S}_1\rket{m_1}={\wtilde S}_2\rket{m_1}={\wtilde S}_3\rket{m_1}=0 \ , 
\label{4-3a}\\
& &{\wtilde S}_2^1\rket{m_1}={\wtilde S}_3^1\rket{m_1}={\wtilde S}_3^2\rket{m_1}=0 \ , 
\label{4-3b}\\
& &{\wtilde S}_1^1\rket{m_1}=-2\sigma\rket{m_1}\ , \qquad
{\wtilde S}_2^2\rket{m_1}=-2\sigma_0\rket{m_1}\ , \qquad
{\wtilde S}_3^3\rket{m_1}=-2\sigma'_0\rket{m_1} \ , \quad
\label{4-3c}\\
& &\sigma'_0=\sigma_0 \ . 
\label{4-3d}
\eeq
\esub
For the convenience of later discussion, we will use $n_0$ and $n$, which are the eigenvalues of 
the quark-number operators ${\wtilde N}_1$, ${\wtilde N}_2$ and 
${\wtilde N}_3$ for $\rket{m_1}$: 
\beq\label{4-4}
{\wtilde N}_1\rket{m_1}=n\rket{m_1}\ , \qquad
{\wtilde N}_2\rket{m_1}={\wtilde N}_3\rket{m_1}=n_0\rket{m_1}\ . 
\eeq
The relation between $(\sigma_0,\sigma)$ and $(n_0, n)$ is given by 
\beq\label{4-5}
\sigma_0=\frac{1}{2}(2j_s+1)-\frac{1}{2}(n_0+n)\ , \qquad
\sigma=\frac{1}{2}(2j_s+1)-n_0\ .
\eeq
Of course, the above expression is also valid in the Schwinger boson realization. 
The form (\ref{3-1}) tells that $\sigma \geq \sigma_0$ and we have 
\beq\label{4-6}
n \geq n_0 \ .
\eeq
The relations (\ref{4-3})$-$(\ref{4-5}) give $\rket{m_1}$ in the following form:
\beq\label{4-7}
\rket{m_1}=\left(\prod_{p=1}^{n-n_0}{\tilde c}_{i\mu_p}^*\right)\left(\prod_{q=1}^{n_0}{\wtilde D}_{\mu'_q}^*\right)\rket{0} \ .
\eeq
Here, ${\wtilde D}_{\mu}^*$ denotes 
\beq\label{4-8}
{\wtilde D}_{\mu}^*=\prod_{i=1}^3{\tilde c}_{i\mu}^* \ .
\eeq
Any of $(\mu_1,\cdots , \mu_{n-n_0},\mu'_1,\cdots ,\mu'_{n_0})$ takes the value between 
$(-j_s)$ to $(+j_s)$ in agreement with the Pauli-principle. 
Therefore,
we have the following state which corresponds to $\dket{\lambda\rho\sigma_0\sigma}$: 
\beq\label{4-9}
\dket{\lambda\rho\sigma_0\sigma}
\rightarrow 
\left({\wtilde S}^3\right)^{2\lambda}\left({\wtilde S}^4(1)\right)^{2\rho}
\left(\prod_{p=1}^{n-n_0}{\tilde c}_{1\mu_p}^*\right)\left(\prod_{q=1}^{n_0}{\wtilde D}_{\mu'_q}^*\right)\rket{0} \ .
\eeq

We rewrote the state (\ref{3-2}) to the form (\ref{3-4}). 
As was already mentioned, the form (\ref{3-4}) is quite suitable for understanding 
the structure of many-quark system. 
We will rewrite the state (\ref{4-9}), following this idea. 
For this task, it may be enough to rewrite the part 
$({\wtilde S}^4(1))^{2\rho}(\prod_{p=1}^{n-n_0}{\tilde c}_{1\mu_p}^*)(\prod_{q=1}^{n_0}{\wtilde D}_{\mu'_q}^*)\rket{0}$ 
to the form suitable for our discussion. 
For this rewriting, we introduce the operator ${\wtilde B}_{\mu}^*$ in the form
\beq\label{4-10}
{\wtilde B}_{\mu}^*=[\ {\wtilde S}^4(1)\ , \ {\tilde c}_{1\mu}^*\ ]\ . 
\eeq
With the use of the explicit form of ${\wtilde S}^4(1)$, ${\wtilde B}_{\mu}^*$ is 
obtained in the following form:
\beq\label{4-11}
{\wtilde B}_{\mu}^*=\sum_{i=1}^3{\wtilde S}^i{\tilde c}_{i\mu}^*\ . 
\eeq
The form (\ref{4-11}) tells that, in spite of the commutator $[{\wtilde S}^4(1)\ , \ {\tilde c}_{1\mu}^*]$ 
which is related to the color $i=1$, the result $\sum_i{\wtilde S}^i{\tilde c}_{i\mu}^*$ 
is color-symmetric. 
The operator ${\wtilde D}_{\mu}^*$ given in the relation (\ref{4-8}) 
is also color-symmetric. 
The operators ${\wtilde B}_{\mu}^*$ and ${\wtilde D}_{\mu}^*$ carry three quarks 
and satisfy the relations 
\bsub\label{4-12}
\beq
& &[\ {\wtilde S}^4(1)\ , \ {\wtilde B}_{\mu}^*\ ]=0 \ , 
\label{4-12a}\\
& &[\ {\wtilde S}^4(1)\ , \ {\wtilde D}_{\mu}^*\ ]=0 \ . 
\label{4-12b}
\eeq
The definition (\ref{4-11}) supports the following anti-commutation relation: 
\beq
\{\ {\tilde c}_{1\mu}^*\ , \ {\wtilde B}_{\mu'}^*\ \}=0\ , \qquad
\{\ {\wtilde B}_{\mu}^*\ , \ {\wtilde B}_{\mu'}^*\ \}=0 \ . 
\label{4-12c} 
\eeq
\esub
Further, we have 
\beq\label{4-13}
{\wtilde S}^4(1)\rket{0}=0\ .
\eeq
If we notice the relations (\ref{4-12a}) and (\ref{4-13}), we have 
\beq\label{4-14}
& &\left({\wtilde S}^4(1)\right)^{2\rho}\left(\prod_{p=1}^{n-n_0}{\tilde c}_{1\mu_p}^*\right)\left(\prod_{q=1}^{n_0}{\wtilde D}_{\mu'_q}^*\right)\rket{0}
\nonumber\\
&=&\left({\vec S}^4(1)\right)^{2\rho}\left(\prod_{p=1}^{n-n_0}{\tilde c}_{1\mu_p}^*\right)\left(\prod_{q=1}^{n_0}{\wtilde D}_{\mu'_q}^*\right)\rket{0}
\ .
\eeq
Here, $({\vec S}^4(1))^{2\rho}(\prod_{p=1}^{n-n_0}{\tilde c}_{1\mu_p}^*)$ denotes the multi-commutator 
defined in the relation (\ref{a1}). 
Let the quantities appearing in Appendix read the following: 
\beq\label{4-15}
{\wtilde O}\rightarrow {\wtilde S}^4(1)\ , \quad
{\tilde c}_{p}^*\rightarrow {\tilde c}_{1\mu_p}^*\ , \quad
{\wtilde B}_{p}^*\rightarrow {\wtilde B}_{\mu_p}^*\ , \quad
L\rightarrow n-n_0 \ , \quad 
M=2\rho \ .
\eeq
Then, the relation (\ref{4-14}) can be expressed as 
\beq\label{4-16}
& &\left({\wtilde S}^4(1)\right)^{2\rho}\left(\prod_{p=1}^{n-n_0}{\tilde c}_{1\mu_p}^*\right)\left(\prod_{q=1}^{n_0}{\wtilde D}_{\mu'_q}^*\right)\rket{0}
\nonumber\\
&=&\frac{1}{(n-n_0-2\rho)!}\sum_P(-)^P P\left(\prod_{p=2\rho+1}^{n-n_0}{\tilde c}_{1\mu_p}^*\prod_{p=1}^{2\rho}{\wtilde B}_{\mu_p}^*\right)
\cdot\left(\prod_{q=1}^{n_0}{\wtilde D}_{\mu'_q}^*\right)\rket{0} \ . 
\eeq
Finally, we obtain the following result: 
\beq\label{4-17}
& &\dket{\lambda\rho\sigma_0\sigma}\rightarrow 
\frac{1}{(n-n_0-2\rho)!}\left({\wtilde S}^3\right)^{2\lambda}
\sum_P(-)^P P\left(\prod_{p=2\rho+1}^{n-n_0}{\tilde c}_{1\mu_p}^*\prod_{p=1}^{2\rho}{\wtilde B}_{\mu_p}^*\right)
\cdot\left(\prod_{q=1}^{n_0}{\wtilde D}_{\mu'_q}^*\right)\rket{0}\nonumber\\
& &
\eeq
Obviously, the state (\ref{4-17}) depends on the colors 3 and 1 and it is 
not color-symmetric. 
Therefore, with the help of the permutation, the state (\ref{4-17}) must be 
symmetrized by the same procedure as the one described in the result (\ref{3-8}). 
Then, we obtain the same results as those in (\ref{3-8})$-$(\ref{3-11}).

Finally, we consider the building blocks of the color-symmetrized version of the 
state (\ref{4-17}): 
${\wtilde S}^k$, ${\tilde c}_{km}^*$, ${\wtilde B}_m^*$ and ${\wtilde D}_m^*$. 
These operators satisfy the relations 
\bsub\label{4-18}
\beq
& &[\ {\wtilde S}_i^j\ , \ {\wtilde S}^k\ ]=\delta_{ij}{\wtilde S}^k+\delta_{jk}{\wtilde S}^i \ , 
\label{4-18a}\\
& &[\ {\wtilde S}_i^j\ , \ {\tilde c}_{km}^*\ ]=\delta_{ij}{\tilde c}_{km}^*-\delta_{ik}{\tilde c}_{jm}^* \ , 
\label{4-18b}
\eeq
\esub
\vspace{-0.6cm}
\bsub\label{4-19}
\beq
& &[\ {\wtilde S}_i^j\ , \ {\wtilde B}_{m}^*\ ]=2\delta_{ij}{\wtilde B}_m^* \ , \qquad\quad
\label{4-19a}\\
& &[\ {\wtilde S}_i^j\ , \ {\wtilde D}_{m}^*\ ]=2\delta_{ij}{\wtilde D}_m^* \ . 
\label{4-19b}
\eeq
\esub
From the definition of ${\wtilde S}^k$ and ${\tilde c}_{km}^*$ and the relation (\ref{4-18}), 
we see that these operators create a quark-pair and a single-quark, respectively, 
and not color-singlet operators. 
On the other hand, ${\wtilde B}_m^*$ and ${\wtilde D}_m^*$ create 
quark-triplet and they are color-singlets. 
But, concerning their role in the present model, both are different from one another. 
The minimum weight state is given by the relation (\ref{4-7}), which consist of 
the single-quark and the quark-triplet operators ${\tilde c}_{1m}^*$ and ${\wtilde D}_m^*$. 
The operator ${\tilde c}_{1m}^*$ is transformed through ${\wtilde S}^4(1)$, but ${\wtilde D}_m^*$ 
is not affected by ${\wtilde S}^4(1)$. 
This can be seen in the relations (\ref{4-10}) and (\ref{4-12b}). 
Therefore, if the system under investigation can be treated in the framework of a single 
irreducible representation, the minimum weight state is unchanged and ${\wtilde B}_m^*$ describes the system, 
alone.
If the description of the system requires at least two irreducible representations 
$(n_0'\neq n_0$), not only ${\wtilde B}_m^*$ but also ${\wtilde D}_m^*$ 
are needed for the description. 
It may be interesting to compare the relations (\ref{3-7}) and (\ref{4-18}) with each other. 
The operators ${\wtilde S}^k$ and ${\tilde c}_{km}^*$ correspond to 
${\hat S}^k$ and ${\hat q}^k$, respectively. 
Judging from the above-mentioned role, ${\wtilde B}_m^*$ corresponds to ${\hat B}^*$. 
In the Schwinger boson realization, we cannot find any operator which corresponds to 
${\wtilde D}_m^*$. 
The part $({\hat b}^*)^{2\sigma_0}\ket{0}$ in the state (\ref{3-2}) does not change 
under the operation of ${\hat S}^4(1)$ given by (\ref{3-3}) on the state 
$({\hat b}^*)^{2\sigma_0}\ket{0}$. 
Further, we have ${\hat N}_i({\hat b}^*)^{2\sigma_0}\ket{0}=n_0({\hat b}^*)^{2\sigma_0}\ket{0}$. 
Therefore, $({\hat b}^*)^{2\sigma_0}\ket{0}$ plays the same role as the state 
$\prod_{q=1}^{n_0}{\wtilde D}_{\mu'_q}^*\rket{0}$.

\section{General form of the $su(4)$-algebraic many-quark model}

With the aim of investigating the variations of the expression (\ref{2-1}), 
we formulate the $su(4)$-algebraic model in a rather general scheme. 
In this section, we treat many-quark system confined in one 
single-particle level which consists of $2\Omega_a$ single-particle states 
in each color. 
Each single-particle state is specified by a quantum number $\alpha$. 
As for $\alpha$, its range of values is given by
\bsub\label{5-1}
\beq
& &\alpha=\pm\frac{1}{2}\ , \ \pm \frac{3}{2} \ , \ \cdots ,\ \pm\left(\Omega_a-\frac{1}{2}\right)\ , \quad
(2\Omega_a:{\rm even\ number})
\label{5-1a}\\
& &\alpha=0\ , \ \pm 1\ , \ \pm 2 \ , \ \cdots ,\ \pm\left(\Omega_a-\frac{1}{2}\right)\ , \quad
(2\Omega_a:{\rm odd\ number})
\label{5-1b}
\eeq
\esub
The relation (\ref{5-1}) tells that, except $\alpha=0$ in the relation (\ref{5-1b}), 
as a partner of $\alpha$, we can choose another single-particle, $-\alpha$, which 
is denoted as ${\tilde \alpha}$. 
In the case $\alpha=0$, the partner of $\alpha$, ${\tilde \alpha}$, is the state 
$\alpha$ it-self, i.e., ${\tilde \alpha}=\alpha$.

Under the above arrangement and with the use of real function $a(\alpha)$, we 
define the following fifteen operators:
\bsub\label{5-2}
\beq
& &{\wtilde S}^1=\sum_{\alpha}a(\alpha){\tilde c}_{2\alpha}^*{\tilde c}_{3{\tilde \alpha}}^* \ , \quad
{\wtilde S}^2=\sum_{\alpha}a(\alpha){\tilde c}_{3\alpha}^*{\tilde c}_{1{\tilde \alpha}}^* \ , \quad
{\wtilde S}^3=\sum_{\alpha}a(\alpha){\tilde c}_{1\alpha}^*{\tilde c}_{2{\tilde \alpha}}^* \ , 
\nonumber\\
& &{\wtilde S}_1=({\wtilde S}^1)^*\ , \qquad 
{\wtilde S}_2=({\wtilde S}^2)^*\ , \qquad
{\wtilde S}_3=({\wtilde S}^3)^* \ , 
\label{5-2a}\\
& &{\wtilde S}_1^2=-\sum_{\alpha}{\tilde c}_{2\alpha}^*{\tilde c}_{1{\alpha}} \ , \quad
{\wtilde S}_1^3=-\sum_{\alpha}{\tilde c}_{3\alpha}^*{\tilde c}_{1{\alpha}} \ , \quad
{\wtilde S}_2^3=-\sum_{\alpha}{\tilde c}_{3\alpha}^*{\tilde c}_{2{\alpha}} \ , 
\nonumber\\
& &{\wtilde S}_2^1=({\wtilde S}_1^2)^*\ , \qquad 
{\wtilde S}_3^1=({\wtilde S}_1^3)^*\ , \qquad
{\wtilde S}_3^2=({\wtilde S}_2^3)^* \ , 
\label{5-2b}\\
& &{\wtilde S}_1^1=\sum_{\alpha}({\tilde c}_{2\alpha}^*{\tilde c}_{2\alpha}
+{\tilde c}_{3\alpha}^*{\tilde c}_{3\alpha}-1)\ , \quad
{\wtilde S}_2^2=\sum_{\alpha}({\tilde c}_{3\alpha}^*{\tilde c}_{3\alpha}
+{\tilde c}_{1\alpha}^*{\tilde c}_{1\alpha}-1)\ , \nonumber\\
& &{\wtilde S}_3^3=\sum_{\alpha}({\tilde c}_{1\alpha}^*{\tilde c}_{1\alpha}
+{\tilde c}_{2\alpha}^*{\tilde c}_{2\alpha}-1)\ . 
\label{5-2c}
\eeq
\esub
Let $a(\alpha)$ and $a({\tilde \alpha})$ obey the condition 
\bsub\label{5-3}
\beq
a(\alpha)^2=1\ , \qquad a({\tilde \alpha})=a(\alpha)\ . 
\label{5-3a}
\eeq
If $a(\alpha)$ does not satisfy the condition (\ref{5-3a}), $a(\alpha)$ should vanish:
\beq
a(\alpha)=0\ . 
\label{5-3b}
\eeq
\esub
We can prove that the above fifteen operators form the $su(4)$-algebra, that is, 
they satisfy the commutation relation (\ref{2-2a}). 
If $a(\alpha)$ is changed in the frame of the condition (\ref{5-3}), we obtain various expressions for 
the $su(4)$-algebra.

Even if $a(\alpha)$ is reasonably chosen under physical interpretation, we encounter 
some cases which do not satisfy the condition (\ref{5-3}). 
In these cases, we supplement $\alpha$ with new quantum number $\beta$ and 
make the following replacement in the expression (\ref{5-2}): 
\beq\label{5-4}
\alpha \rightarrow \alpha\beta\ , \qquad 
a(\alpha)\rightarrow a(\alpha)b(\beta)\ , \qquad
2\Omega_a\rightarrow 4\Omega_a\Omega_b\ .
\eeq
We obtain the expression satisfying the $su(4)$-algebra, if $a(\alpha)b(\beta)$ 
obey the condition 
\beq\label{5-5}
(a(\alpha)b(\beta))^2=1\ . \qquad
a({\tilde \alpha})b({\tilde \beta})=a(\alpha)b(\beta)\ . 
\eeq
Of course, $\beta$ is divided into two cases which are similar to those in 
(\ref{5-1a}) and (\ref{5-1b}). 
If the set of the quantum number $(\alpha,\beta)$ still does not satisfy 
the condition (\ref{5-5}), we proceed with the same 
task as before and make the following replacement: 
\beq\label{5-6}
\alpha \rightarrow \alpha\beta\gamma\ , \qquad 
a(\alpha)\rightarrow a(\alpha)b(\beta)c(\gamma)\ , \qquad
2\Omega_a\rightarrow 8\Omega_a\Omega_b\Omega_c\ .
\eeq
Of course, $a(\alpha)b(\beta)c(\gamma)$ should satisfy 
\beq\label{5-7}
(a(\alpha)b(\beta)c(\gamma))^2=1\ . \qquad
a({\tilde \alpha})b({\tilde \beta})c({\tilde \gamma})=a(\alpha)b(\beta)c(\gamma)\ . 
\eeq
Here, $\gamma$ is divided into two cases which are similar to those in 
(\ref{5-1a}) and (\ref{5-1b}).

The expression of the $su(4)$-generators (\ref{5-2}) suggests us that the idea developed in 
last section is available without any modification. 
The minimum weight state $\rket{m_1}$ is given in the form 
\beq
& &\rket{m_1}=\left(\prod_{p=1}^{n-n_0}{\tilde c}_{1\alpha_p}^*\right)
\left(\prod_{q=1}^{n_0}{\wtilde D}_{\alpha_q'}^*\right)\rket{0}\ , 
\label{5-8}\\
& &{\wtilde D}_{\alpha}^*=\prod_{i=1}^3{\tilde c}_{i\alpha}^*\ . 
\label{5-9}
\eeq
The above forms come from the relations (\ref{4-7}) and (\ref{4-8}). 
The operator ${\wtilde B}_{\alpha}^*$ is given in the form 
\beq\label{5-10}
{\wtilde B}_{\alpha}^*=[\ {\wtilde S}^4(1)\ , \ {\tilde c}_{1\alpha}^*\ ]=
\sum_{i=1}^3{\wtilde S}^i{\tilde c}_{i\alpha}^*\ .
\eeq
Of course, ${\wtilde B}_{\alpha}^*$ and ${\wtilde D}_{\alpha}^*$ satisfy 
\bsub\label{5-11}
\beq
& &[\ {\wtilde S}^4(1)\ , \ {\wtilde B}_{\alpha}^*\ ]=0 \ , 
\label{5-11a}\\
& &[\ {\wtilde S}^4(1)\ , \ {\wtilde D}_{\alpha}^*\ ]=0 \ , 
\label{5-11b}\\
& &\{\ {\tilde c}_{1\alpha}^*\ , \ {\wtilde B}_{\alpha'}^*\ \}=0\ , \qquad
\{\ {\wtilde B}_{\alpha}^*\ , \ {\wtilde B}_{\alpha'}^*\ \}=0\ .
\label{5-11c}
\eeq
\esub
Through the same process as that in \S 4, 
we have the following form: 
\beq\label{5-12}
\dket{\lambda\rho\sigma_0\sigma}
&\rightarrow&
\frac{1}{(n-n_0-2\rho)!}\left({\wtilde S}^3\right)^{2\lambda}\nonumber\\
& &\times
\sum_P(-)^P P\left(\prod_{p=2\rho+1}^{n-n_0}{\tilde c}_{1\alpha_p}^*\prod_{p=1}^{2\rho}{\wtilde B}_{\alpha_p}^*
\right)\cdot\left(\prod_{q=1}^{n_0}{\wtilde D}_{\alpha_q'}^*\right)\rket{0} \ .
\eeq
It may be not necessary to give the explanation of the notations. 
In the case where the condition (\ref{5-3}) does not hold, it may be 
enough to make the replacement (\ref{5-4}) or (\ref{5-6}) to the 
relations (\ref{5-8})$-$(\ref{5-12}). 
Judging from the above treatment, all the relations derived in \S 4 are 
valid in general, if the quantum number specifying the single-particle 
state is changed from $m$ to $\alpha$. 
Therefore, the relations (\ref{4-18}) and (\ref{4-19}) which characterized the 
building blocks are valid if $m$ is replaced by $\alpha$ and the argument below 
the relations (\ref{4-18}) and (\ref{4-19}) is valid.

\section{Discussions and some examples}

Main task of this section is to apply the general form presented in \S 5 to 
certain concrete cases. 
For this aim, we return to the form given in \S 2. 
Let us put $\alpha=m$ and $\Omega_a-1/2=j_s$ in the relation (\ref{5-1a}). 
Then, the expression (\ref{2-1}) is obtained by putting $a(\alpha)=1$ for 
all $\alpha$ in the relation (\ref{5-2}) and, certainly, this case satisfies 
the condition (\ref{5-3}). 
In this sense, the expression (\ref{2-1}) is one of the examples of the general 
form (\ref{5-2}). 
Of course, we have $2\Omega_a=2j_s+1$. 
If we regard $j_s$ as the angular momentum specifying the single-particle 
level in the spherical $j$-$j$ coupling shell model, 
the quantum number $m$ becomes the projection to $z$-axis. 
In the case $a(\alpha)=1$, ${\wtilde S}^i$ does not 
indicate the quark-pair coupled to the angular momentum $J=0$. 
If we adopt $a(\alpha)=(-)^{j_s-m}$, ${\wtilde S}^i$ becomes 
the quark-pair coupled to $J=0$, because of the Clebsch-Gordan coefficient 
$\langle j_smj_s-m \ket{00}=(-)^{j_s-m}\cdot 1/\sqrt{2j_s+1}$. 
However, in this case, we have $a({\tilde \alpha})=(-)^{j_s+m}=-a({\alpha})$ and 
in the frame of $a(\alpha)=(-)^{j_s-m}$, the $su(4)$-algebra cannot be expected. 
In order to overcome this discrepancy, inevitably, we must the quantum number $\beta$. 
The spin and the isospin are the properties of the quark except the color. 
Since $j_s$ is a half-integer, the effect of the spin is already included in $\alpha$ and 
we have an idea to introduce the isospin as $\beta$.

Under the above consideration, let us introduce the isospin to our model. 
The single-particle state is specified by two quantum number $m$ and $\tau$ $(=\pm 1/2)$, 
i.e., $b(\beta)=(-)^{\frac{1}{2}-\tau}$ and $2\Omega_b=2$ and $a(\alpha)b(\beta)$ is 
expressed as 
\beq\label{6-1}
a(\alpha)b(\beta)=(-)^{j_s-m}(-)^{\frac{1}{2}-\tau}\ , \qquad
(a(\alpha)b(\beta))^2=1\ . 
\eeq
Certainly, we have
\beq\label{6-2}
a({\tilde \alpha})b({\tilde \beta})=
(-)^{j_s+m}(-)^{\frac{1}{2}+\tau}=a(\alpha)b(\beta) \ . 
\eeq
The relation (\ref{6-1}) tells us that we are considering the quark-pair with 
$J=0$ and $T=0$. 
By substituting the relation (\ref{6-1}) into the form (\ref{5-2}), we obtain the expression of the 
$su(4)$-generators in the present case. 
For example, we have ${\wtilde S}^1=\sum_{m\tau}(-)^{j_s-m}(-)^{\frac{1}{2}-\tau}
{\tilde c}_{2m\tau}^*{\tilde c}_{3{\tilde m}{\tilde \tau}}^*$ and 
${\wtilde S}_1^1=\sum_{m\tau}({\tilde c}_{2m\tau}^*{\tilde c}_{2m\tau}+{\tilde c}_{3m\tau}^*{\tilde c}_{3m\tau})
-2(2j_s+1)$. 
The quark-triplets can be expressed as 
\beq\label{6-3}
{\wtilde B}_{m\tau}^*=\sum_{i=1}^3{\wtilde S}^i{\tilde c}_{im\tau}^*\ , \qquad
{\wtilde D}_{m\tau}^*=\prod_{i=1}^3{\tilde c}_{im\tau}^*\ .
\eeq
The form (\ref{6-3}) is nothing but the form proposed in the original Bonn model, 
in which the former and the latter represent ``nucleon" and $\Delta$-excitation. 
In the introductory part of \S 5, 
we showed two cases. 
The above example is based on the case (\ref{5-1a}).

Next, we consider the case (\ref{5-1b}). 
In parallel with $(j_s, m)$, we introduce a set of the integers $(l_s,\mu)$. 
For a given $l_s$, $\mu$ takes the value $\mu=0,\ \pm1,\ \pm 2,\cdots ,\ \pm l_s$ and 
we put $\alpha=\mu$ and $\Omega_a-1/2=l_s$ in the 
relation (\ref{5-1b}). 
In the same idea as that in the case $(j_s, m)$, we regard $(l_s, \mu)$ as the 
angular momentum (orbital). 
The function $a(\alpha)=(-)^{l_s-\mu}$ satisfies $a({\tilde \alpha})=(-)^{l_s+\mu}=a(\alpha)$ 
and, then, different from the previous case, we obtain the $su(4)$-algebra. 
However, in the frame of $(l_s, \mu)$, only the orbital angular momentum is taken into account. 
Therefore, for the problem of the quark, we must introduce the degrees of freedom 
related to the spin and the isospin. 
The single-particle state is specified by three quantum numbers $\mu$, $\sigma(=\pm 1/2)$ and 
$\tau(=\pm 1/2)$, i.e., 
$b(\beta)=(-)^{\frac{1}{2}-\sigma}$, 
$c(\gamma)=(-)^{\frac{1}{2}-\tau}$, $2\Omega_b=2$ and $2\Omega_c=2$. 
Then, $a(\alpha)b(\beta)c(\gamma)$ is expressed as 
\beq\label{6-4}
a(\alpha)b(\beta)c(\gamma)=(-)^{l_s-\mu}(-)^{\frac{1}{2}-\sigma}(-)^{\frac{1}{2}-\tau}\ , 
\qquad
(a(\alpha)b(\beta)c(\gamma))^2=1\ .
\eeq
From the following relation, we can expect the $su(4)$-algebra: 
\beq\label{6-5}
a({\tilde \alpha})b({\tilde \beta})c({\tilde \gamma})=(-)^{l_s+\mu}(-)^{\frac{1}{2}+\sigma}(-)^{\frac{1}{2}+\tau}
=a(\alpha)b(\beta)c(\gamma)\ .
\eeq
For instance, we have the relations ${\wtilde S}^1=\sum_{\mu\sigma\tau}(-)^{l_s-\mu}(-)^{\frac{1}{2}-\sigma}(-)^{\frac{1}{2}-\tau}
{\tilde c}_{2\mu\sigma\tau}^*{\tilde c}_{3{\tilde \mu}{\tilde \sigma}{\tilde \tau}}^*$ and\break
${\wtilde S}_1^1=\sum_{\mu\sigma\tau}({\tilde c}_{2\mu\sigma\tau}^*{\tilde c}_{2\mu\sigma\tau}
+{\tilde c}_{3\mu\sigma\tau}^*{\tilde c}_{3\mu\sigma\tau})-4(2l_s+1)$. 
In the same manner as the previous case, the quark-triplets in the present case are given 
in the form
\beq\label{6-6}
{\wtilde B}_{\mu\sigma\tau}^*=\sum_{i=1}^3{\wtilde S}^i{\tilde c}_{i\mu\sigma\tau}^* \ , \qquad
{\wtilde D}_{\mu\sigma\tau}^*=\prod_{i=1}^3{\tilde c}_{i\mu\sigma\tau}^* \ . 
\eeq
The former and the latter denote ``nucleon" and $\Delta$-excitation. 
In contrast with the case (\ref{5-1a}), the case (\ref{5-1b}) is based on the 
spherical $L$-$S$ coupling shell model.

As is clear from the above argument, we have two forms for the ``nucleon". 
The forms (\ref{6-3}) and (\ref{6-6}) are based on the spherical $j$-$j$ coupling 
and $L$-$S$ coupling shell models, respectively. 
In subsequent paper, we will discuss these two forms and some problems related to them.

Concerning the condition (\ref{5-1b}), we have discussed the case $a(\alpha)=(-)^{l_s-\mu}$. 
The case $a(\alpha)=1$ for all $\alpha$ gives us the $su(4)$-algebra and this case 
can be regarded as the correspondence of the condition (\ref{5-1a}) with $a(\alpha)=1$. 
As a case which cannot be found in the condition (\ref{5-1a}), 
we consider the $a(\alpha)=1$ with $2\Omega_a=1$. 
This is nothing but the case $\alpha=0$ which consists of $({\tilde c}_{i0}^* , {\tilde c}_{i0})$ 
for $i=1$, 2 and 3. 
But, if limited to one single-particle level with $\alpha=0$, the model 
is too simple to intend to investigate. 
Then, we enlarge the number of the single-particle level 
from one to the plural and each level is specified as $\nu$ $(\nu=0,\ 1,\ 2,\cdots ,\ \nu_0)$. 
Therefore, the single-particle state is specified by $i\nu$: $({\tilde c}_{i\nu}^*, {\tilde c}_{i\nu})$.

The $su(4)$-generators for $\nu$-th single-particle level are 
expressed as follows:
\bsub\label{6-7}
\beq
& &{\wtilde S}^1(\nu)={\tilde c}_{2\nu}^*{\tilde c}_{3\nu}^*\ , \qquad
{\wtilde S}^2(\nu)={\tilde c}_{3\nu}^*{\tilde c}_{1\nu}^*\ , \qquad
{\wtilde S}^3(\nu)={\tilde c}_{1\nu}^*{\tilde c}_{2\nu}^*\ , \nonumber\\
& &{\wtilde S}_1(\nu)=({\wtilde S}^1(\nu))^*\ , \qquad
{\wtilde S}_2(\nu)=({\wtilde S}^2(\nu))^*\ , \qquad
{\wtilde S}_3(\nu)=({\wtilde S}^3(\nu))^*\ , 
\label{6-7a}\\
& &{\wtilde S}_1^2(\nu)=-{\tilde c}_{2\nu}^*{\tilde c}_{1\nu}\ , \qquad
{\wtilde S}_1^3(\nu)=-{\tilde c}_{3\nu}^*{\tilde c}_{1\nu}\ , \qquad
{\wtilde S}_2^3(\nu)=-{\tilde c}_{3\nu}^*{\tilde c}_{2\nu}\ , \nonumber\\
& &{\wtilde S}_2^1(\nu)=({\wtilde S}_1^2(\nu))^*\ , \qquad
{\wtilde S}_3^1(\nu)=({\wtilde S}_1^3(\nu))^*\ , \qquad
{\wtilde S}_3^2(\nu)=({\wtilde S}_2^3(\nu))^*\ , 
\label{6-7b}\\
& &{\wtilde S}_1^1(\nu)={\tilde c}_{2\nu}^*{\tilde c}_{2\nu}
+{\tilde c}_{3\nu}^*{\tilde c}_{3\nu}-1\ , \qquad
{\wtilde S}_2^2(\nu)={\tilde c}_{3\nu}^*{\tilde c}_{3\nu}
+{\tilde c}_{1\nu}^*{\tilde c}_{1\nu}-1\ , \nonumber\\
& &{\wtilde S}_3^3(\nu)={\tilde c}_{1\nu}^*{\tilde c}_{1\nu}
+{\tilde c}_{2\nu}^*{\tilde c}_{2\nu}-1\ . 
\label{6-7c}
\eeq
\esub
Further, we have the relation for $\nu'\neq \nu$
\beq\label{6-8}
[\ {\rm any\ of\ the}\ \nu'\hbox{\rm -th\ generators}\ , \ {\rm any\ of\ the}\ \nu\hbox{\rm -th\ generators}\ ]=0 \ .
\eeq
Fermion number operator in the $\nu$-th level, 
${\wtilde N}(\nu)$, is expresses as 
\bsub\label{6-9}
\beq
{\wtilde N}(\nu)=\sum_i{\wtilde N}_i(\nu)\ , \qquad
{\wtilde N}_i(\nu)={\tilde c}_{i\nu}^*{\tilde c}_{i\nu}
=\frac{1}{2}\left(\sum_j{\wtilde S}_j^j(\nu)+1\right)-{\wtilde S}_{i}^i(\nu) \ . 
\label{6-9a}
\eeq
Fermion number operator in the color $i$ and the total fermion number 
operator are expressed as 
\beq
& &{\wtilde N}_i=\sum_{\nu}{\wtilde N}_i(\nu)={\wtilde N}-\sum_{\nu}{\wtilde S}_i^i(\nu)-\nu \ , 
\label{6-9b}\\
& &{\wtilde N}=\sum_{\nu}{\wtilde N}(\nu)\ .
\label{6-9c}
\eeq
\esub
The orthogonal set in the $\nu$-th subspace is given as follows: 
\beq\label{6-10}
& &\rket{0}\ , \quad {\tilde c}_{2\nu}^*{\tilde c}_{3\nu}^*\rket{0}\ , \quad
{\tilde c}_{3\nu}^*{\tilde c}_{1\nu}^*\rket{0}\ , \quad
{\tilde c}_{1\nu}^*{\tilde c}_{2\nu}^*\rket{0}\ , \nonumber\\
& &{\tilde c}_{1\nu}^*\rket{0}\ , \quad 
{\tilde c}_{2\nu}^*\rket{0}\ , \quad
{\tilde c}_{3\nu}^*\rket{0}\ , \quad
{\tilde c}_{1\nu}^*{\tilde c}_{2\nu}^*{\tilde c}_{3\nu}^*\rket{0}\ .
\eeq
The vacuum $\rket{0}$ is common to all subspace.

The Hamiltonian adopted in this model, for example, is expressed in the form 
\beq\label{6-11}
{\wtilde H}=\sum_{\nu}\epsilon_{\nu}{\wtilde N}(\nu)-G\sum_i\left(\sum_{\nu}{\wtilde S}^i(\nu)\right)
\left(\sum_\nu{\wtilde S}_i(\nu)\right) \ . \quad
(G>0)
\eeq
Here, $\epsilon_{\nu}$ and $G$ denote the single-particle energy of the $\nu$-th level 
and the interaction strength, respectively. 
The order of the level is given as $\epsilon_1 \leq \epsilon_2 \leq \cdots \leq \epsilon_{\nu}$. 
This model is a kind of $su(4)\otimes su(4)\otimes \cdots \otimes su(4)$-algebraic model. 
If $\epsilon_1=\epsilon_2= \cdots =\epsilon_{\nu}(=\epsilon)$, the Hamiltonian (\ref{6-11}) 
reduces to the following:
\beq\label{6-12}
{\wtilde H}=\epsilon{\wtilde N}-G\sum_i{\wtilde S}^i{\wtilde S}_i \ , \qquad
{\wtilde S}^i=\sum_{\nu}{\wtilde S}^i(\nu) \ . 
\eeq
The Hamiltonian (\ref{6-12}) may be equivalent to that of the Bonn model. 
The above model has been investigated by Errea et al.\cite{8} by the 
name of the model for a trapped three-color atom gas. 
If we adopt the Schwinger boson representation, this model may be described in the 
form much simpler than that by Errea et al. 
The conventional and familiar many-body technique can be easily applied. 
Of course, in this case, the Schwinger boson space is constructed by 
the bosons $({\hat a}_i(\nu), {\hat a}_i^*(\nu))$, 
$({\hat b}_i(\nu), {\hat b}_i^*(\nu))$, $({\hat a}(\nu), {\hat a}^*(\nu))$ and 
$({\hat b}(\nu), {\hat b}^*(\nu))$ for $i=1$, 2, 3 and $\nu=0,\ 1,\ 2, \cdots ,\ \nu_0$.

\section{Summary}

In this paper, we have re-formulated the many-quark model with the $su(4)$-algebraic structure developed 
in our previous papers (I) and Refs.\citen{4} and \citen{5}. 
In those papers, we investigated the modified Bonn quark model by using the Schwinger boson realization 
in the boson space and constructed the exact eigenstates together with exact eigenvalues for the energy 
under consideration. 
The eigenstates were constructed by the use of operators with the number of quarks, 1, 2 and 3, 
which were interpreted as the single-quark, the quark-pair and the quark-triplet operator, respectively, 
in the boson space. 
In this paper, these operators have been transcribed into the quark operators in 
the original fermion space. 
Through this transcription, it was clarified that the single-quark operators were the 
quark operators themselves. 
Further, the structure of the quark-triplet operators were also clarified, in which 
the ``nucleon" and ``$\Delta$-excitation" were indicated in the original fermion space.

Furthermore, in this paper, it has been shown that, following the choice of the quantum numbers of the 
original fermion operator besides the color quantum number, 
we can construct the various $su(4)$-algebraic models. 
If the flavor (isospin) quantum number is introduced adding to the total angular momentum $j_s$ 
in a single-particle level, the Bonn quark model can be obtained in the $j$-$j$ coupling scheme. 
When the orbital angular momentum $(l_s,\mu)$, instead of $(j_s,m)$, is treated together with 
the quark spin $(1/2,\pm 1/2)$, the Bonn quark model based on the $L$-$S$ coupling scheme appears. 
Further, if many single-particle levels are introduced, the 
$su(4)\otimes su(4)\otimes \cdots \otimes su(4)$-algebraic model can be constructed, which 
may be known as a model of a trapped three color atom gas in the atomic physics. 
Thus, it is concluded that the many-fermion models with the $su(4)$-algebraic structure 
may be treated widely in our scheme developed in our papers and re-formulated in this paper.

\section*{Acknowledgement}

One of the authors (Y.T.) 
is partially supported by the Grants-in-Aid of the Scientific Research 
(No.23540311) from the Ministry of Education, Culture, Sports, Science and 
Technology in Japan.

\appendix
\section{General formula for deriving the relation (\ref{4-16})}

In this Appendix, we will present a formula for deriving the relation (\ref{4-16}) in a rather 
general form. 
Our present problem is to give an idea how to express the multiple 
commutator of ${\wtilde O}$ for 
${\tilde c}_L^*\cdots {\tilde c}_1^*$ which is defined in the following form: 
\beq\label{a1}
({\vec O})^M({\tilde c}_L^*\cdots {\tilde c}_1^*)=
[\ \underbrace{{\wtilde O}\ , \ [\ {\wtilde O}\ , \cdots ,\ [\ {\wtilde O}}_{M} \ , \ {\tilde c}_L^*\cdots {\tilde c}_1^* \ ] 
\cdots ]\ .
\eeq
The symbol ${\vec O}{\wtilde X}$ denotes the commutator $[\ {\wtilde O}\ , \ {\wtilde X}\ ]$ and 
$({\tilde c}_l^*, {\tilde c}_l; l=1,2,\cdots ,L)$ is a set of fermion operators. 
The operator ${\wtilde O}$ is a function of $({\tilde c}_l^*, {\tilde c}_l; l=1,2,\cdots ,L)$ and 
we require the following condition: 
\beq\label{a2}
[\ {\wtilde O}\ , \ {\wtilde B}_l^*\ ]=0\ , \qquad
\{\ {\tilde c}_l^*\ , \ {\wtilde B}_{l'}^*\ \}=0\ , \qquad 
\{\ {\wtilde B}_l^*\ , \ {\wtilde B}_{l'}^*\ \}=0\ . 
\eeq
Here, ${\wtilde B}_l^*$ is defined as 
\beq\label{a3}
{\wtilde B}_l^*=[\ {\wtilde O}\ , \ {\tilde c}_l^*\ ]\ .
\eeq

First, we notice the identity 
\beq\label{a4}
{\tilde c}_{L}^*\cdots {\tilde c}_1^*=(L!)^{-1}\sum_P(-)^P P({\tilde c}_L^*\cdots {\tilde c}_1^*)\ . 
\eeq
Here, $P$ denotes the permutation 
\beq\label{a5}
P=\left(
\begin{array}{cccc}
1 & 2 & \cdots & L \\
p_1 & p_2 & \cdots & p_L \\
\end{array}
\right)\ , 
\qquad
(-)^P=\left\{
\begin{array}{cl}
+1 & ; {\rm even\ permutation}\\
-1 & ; {\rm odd\ permutation}\\
\end{array}
\right.
\eeq
The simplest case is as follows:
\beq\label{a6}
& &{\vec O}({\tilde c}_L^*\cdots {\tilde c}_1^*)=(L!)^{-1}\sum_P(-)^P P\bigl({\vec O}({\tilde c}_L^*\cdots {\tilde c}_1^*)) \nonumber\\
&=&(L!)^{-1}\sum_P(-)^P P({\wtilde B}_L^*{\tilde c}_{L-1}^*\cdots {\tilde c}_2^*{\tilde c}_1^*
+{\tilde c}_L^*{\wtilde B}_{L-1}^*\cdots {\tilde c}_2^*{\tilde c}_1^*+\cdots
\nonumber\\
& &\qquad\qquad\qquad\qquad
+{\tilde c}_L^*{\tilde c}_{L-1}^*\cdots {\wtilde B}_2^*{\tilde c}_1^*+
{\tilde c}_L^*{\tilde c}_{L-1}^*\cdots {\tilde c}_2^*{\wtilde B}_1^*\bigl)\nonumber\\
&=&(L!)^{-1}L\sum_P(-)^P P({\tilde c}_L^*{\tilde c}_{L-1}^*\cdots {\tilde c}_2^*{\wtilde B}_1^*)\nonumber\\
&=&((L-1)!)^{-1}\sum_P(-)^P P({\tilde c}_L^*{\tilde c}_{L-1}^*\cdots {\tilde c}_2^*{\wtilde B}_1^*)\ .
\eeq
Of course, we used the relation (\ref{a2}) and (\ref{a3}). 
By calculating successively to the higher power for ${\vec O}$, we 
obtain the following form: 
\beq\label{a7}
({\vec O})^M({\tilde c}_L^*\cdots {\tilde c}_1^*)
=((L-M)!)^{-1}\sum_P (-)^P P({\tilde c}_L^*\cdots {\tilde c}_{M+1}^*{\wtilde B}_M^*\cdots {\wtilde B}_1^*)\ . 
\eeq
Especially, for the case $M=L$, we have 
\beq\label{a8}
({\vec O})^L({\tilde c}_L^*\cdots {\tilde c}_1^*)
&=&\sum_P(-)^P P\left({\wtilde B}_L^*\cdots {\wtilde B}_1^*\right)\nonumber\\
&=&L!\left({\wtilde B}_L^*\cdots {\wtilde B}_1^*\right) \ . 
\eeq
We can see that all ${\tilde c}_l^*$ are replaced with the corresponding ${\wtilde B}_l^*$.


\end{document}